\documentclass[twocolumn,aps, prl,showpacs]{revtex4}

\usepackage{graphics}
\usepackage{graphicx}
\usepackage{dcolumn}
\usepackage{bm}
\usepackage{amsmath,amssymb}

\newcommand{\de}{\partial}

\newcommand{\eq}[2]{\begin{equation} \label{#1} #2 \end{equation}}

\begin{document}

\title{Optical analogues of Bloch-Zener oscillations in binary waveguide arrays: wavenumber evolution perspective}
\author{Minh C. Tran $^{1,2,*}$ and Truong X. Tran$^{3,\dag}$}

\affiliation{$^{1}$ Atomic Molecular and Optical Physics Research Group, Science and Technology Advanced Institute, Van Lang University, Ho Chi Minh City, Vietnam\\
 $^{2}$ Faculty of Applied Technology, School of Technology, Van Lang University, Ho Chi Minh City, Vietnam\\
  $^{3}$ Department of Physics, Le Quy Don Technical University, 236 Hoang Quoc Viet street, 10000 Hanoi, Vietnam\\
 $^{*}$ trancongminh@vlu.edu.vn\\
 $\dag$ Corresponding author: tranxtr@gmail.com}
\date{\today}

\begin{abstract}
We study optical analogues of Bloch oscillations and Zener tunneling in binary waveguide arrays (BWAs) with the help of the wavenumber-based approach. We analytically find two very simple laws describing the evolution of the central wavenumbers of beams in BWAs. From these simple laws, we can easily obtain the propagation distances in the analytical form where the beams operate at the Dirac points, and therefore, the Zener tunneling takes place due to the interband transition. We can also easily calculate the distances where beams reach the turning points in their motion. These distances just depend on the strength of the linear potential and the initial wavenumber of input beams. We also show that the nonlinearity of the Kerr type has a detrimental influence on the Bloch-Zener oscillations.
\end{abstract}
\maketitle

\section{I. INTRODUCTION}
\label{Introduction}

Waveguide arrays (WAs) are a remarkable platform to study discrete photonic effects such as discrete diffraction \cite{jones}, discrete optical solitons \cite{christodoulides,kivshar,lederer}, diffractive resonant radiation \cite{tranresonant1}, and discrete solitons self-wavenumber shift \cite{tranresonant2}.

Especially, conventional WAs consisting of identical waveguides have been intensively used for simulating basic nonrelativistic quantum mechanics effects arising from the Schr\"{o}dinger equation such as Bloch oscillations \cite{bloch,blochZener,pertsch,lenz} and Zener tunneling \cite{ghulinyan,trompeter}. This is because one can convert the coupled-mode equations governing the light beam propagation in WAs into the Schr\"{o}dinger equation. However, for simulating relativistic quantum mechanics phenomena emerging from the Dirac equation one needs to use binary waveguide arrays (BWAs) which are a special class of WAs consisting of two different alternating types of waveguides. Thanks to the possibility of converting the coupled-mode equations in BWAs into the Dirac equation, one can use BWAs to mimic many relativistic quantum mechanics phenomena such as {\em Zitterbewegung} \cite{zitterbewegung}, Dirac solitons (DSs) \cite{trandirac2}, Klein tunneling \cite{klein,kleinLonghi,kleinDreisow,tranKT1,tranKT2}, Sauter effect \cite{sauter,tranSauter}, electron-positron pair production and annihilation \cite{longhiPP,dreisowPP,tranPP1,tranPP2}, and topological Jackiw-Rebbi states \cite{jackiw,tranjr3,tranjr4}. Thanks to Jackiw-Rebbi states, the fundamental charge fractionalisation effect was predicted which led to the fractional quantum Hall effect \cite{laughlin}. The Jackiw-Rebbi states are also fundamental in topological insulators \cite{hasan} which have great potential in designing  robust optical circuits \cite{rechtsman}.

The behaviors of electrons in crystal lattices under the action of an external linear potential (or a dc electric field) described by the Schr\"{o}dinger equation were first studied by Bloch \cite{bloch} and Zener \cite{blochZener}. In 1934 Zener \cite{blochZener} predicted that in this case, electron wave packets undergo periodic oscillations called Bloch oscillations (BOs) which are related to the formation of the electronic Wannier-Stark ladder energy spectrum. However, only after more than half a century later, the electronic Wannier-Stark ladders \cite{wannier,mendez} and BOs \cite{feldmann} were experimentally observed with the invention of superlattices made of semiconductors. Because one just needs a coherent wave in a periodic lattice exposed to an external linear potential to observe BOs, other physical settings such as ultracold atoms in accelerated optical lattices \cite{dahan}, photons in a chirped fiber grating \cite{sterke}, photons in periodic dielectric multilayers \cite{sapienza}, and photons in WAs \cite{pertsch,morandotti} have been used to observe BOs.

Zener also supposed that BOs cannot last forever, but are damped by interband transitions. This is called Zener tunneling (ZT) or electrical Zener breakdown. In terms of applications, the ZT is even more important than BOs because, unlike BOs, the ZT can induce a dc current. Thanks to the ZT, it is possible to realize the electrical breakdown in dielectrics with Zener diodes \cite{blochZener}, electrical conduction along nanotubes \cite{bourlon} and through superlattices \cite{sibille}, and tunneling through Josephson junction \cite{ithier}.

In waveguide arrays analyzed in Ref. \cite{trompeter}, one can observe the irreversible decaying of the BOs due to the ZT to higher-order bands. The ZT in Ref. \cite{trompeter} reveals itself as regular outbursts of radiation which escapes at the turning points of the light beam motion at the edge of the Brillouin zone.    However, as in semiconductor superlattices \cite{rotvig}, it is possible to observe a coherent dynamics instead of and irreversible decay when only two bands exist. Indeed, in binary waveguide arrays which possess only two minibands \cite{sukhorukov1} the coherent Bloch-Zener oscillations have been observed in Refs. \cite{breid,longhiZener,dreisowZener} where the ZT shows up as an aperiodic (generally) or periodic beam splitting and recombination of light beam propagation superimposed to the BOs.

In this work, inspired by the earlier achievements in investigating Bloch-Zener oscillations (BZOs) in BWAs, we investigate BZOs by using a new approach based on the evolution of transverse wavenumbers of beams in BWAs. We show that one can analytically get the laws for the evolution of the central wavenumbers of beams in BWAs which are in perfect agreement with simulation results for the beam propagation in BWAs in the linear regime. This provides more physical insights into the BZOs. Thanks to that, we can analytically calculate the distances where the ZT takes place and the distances where the beams reach the turning points in their motion depending just on the initial wavenumber of the beam and the strength of the external linear potential imposed upon BWAs. The paper is organized as follows. In Section II we provide the theoretical background for getting expressions describing the wavenumber evolution in WAs and BWAs in the linear regime in the tight-binding approximation. In Section III we show that Bloch oscillations in conventional WAs can also be interpreted as Bloch-Zener oscillations. In Section IV we study the BZOs in BWAs in the linear regime. In Section V we investigate the influence of the nonlinearity on the BZOs. Finally, in Section VI we summarize our results and finish with concluding remarks.

\section{II. GOVERNING EQUATIONS AND ANALYTICAL EXPRESSIONS FOR THE TRANSVERSE WAVENUMBER OF BEAMS}
\label{WA}

Within the framework of the tight-binding approximation, the beam light evolution in a BWA with a linear refractive index gradient in the transverse direction and with Kerr nonlinearity is governed by the following dimensionless coupled-mode equations \cite{tranKT1,tranKT2}:
\begin{equation} \label{CWCM} 
    i\frac{{da_n }}{{dz}} =  -\alpha n a_{n} - \kappa \left[ {a_{n + 1}  + a_{n - 1} } \right] + \left( { - 1} \right)^n \sigma a_n  - \gamma \left| {a_n } \right|^2 a_n,
\end{equation}
where $a_{n}$ denotes the electric amplitude in the $n$th waveguide with position $n$ discretely running from [$-$(N$-$1)/2,...$-$1, 0, 1...(N$-$1)/2] and $N$ being the total (odd) number of all waveguides in BWAs; $z$ is the variable along the longitudinal axis of BWAs; 2$\sigma$ and $\kappa$ represent, respectively, the propagation mismatch and the coupling coefficient between two neighboring waveguides of BWAs; $\gamma$ represents the nonlinear coefficient of waveguides which can be either positive (for self-focusing Kerr nonlinearity) or negative (for self-defocusing Kerr nonlinearity); and $\alpha$ represents the linear gradient of the waveguide refractive index in the transverse direction of BWAs which can be achieved by transversely imposing the temperature gradient upon BWAs \cite{pertsch}, or by using curved BWAs \cite{dreisowZener}. This linear gradient of the waveguide refractive index mimics the external linear electric potential imposed upon periodic lattices whose role is performed by periodic BWAs. That is the reason why optical phenomena governed by Eqs. (\ref{CWCM}) can simulate the BZOs in solid-state physics. The quantity $[\alpha n - (-1)^{n}\sigma]$ represents the propagation constant of the $n$th waveguide of BWAs. One can normalize variables in Eqs. (\ref{CWCM}) such that both $\kappa$ and $|\gamma|$ are equal to unity. Note also that one can have both the positive and negative values of $\sigma$ for the same BWA just by shifting the waveguide position $n$ of the BWA by \emph{any odd} number. One can convert Eqs. (\ref{CWCM}) into the Dirac equation describing electrons under action of a linear electric potential (or constant electric field) represented by $\alpha n$ \cite{tranKT2,tranSauter}.

\begin{figure}[htb]
  \centering \includegraphics[width=0.45\textwidth]{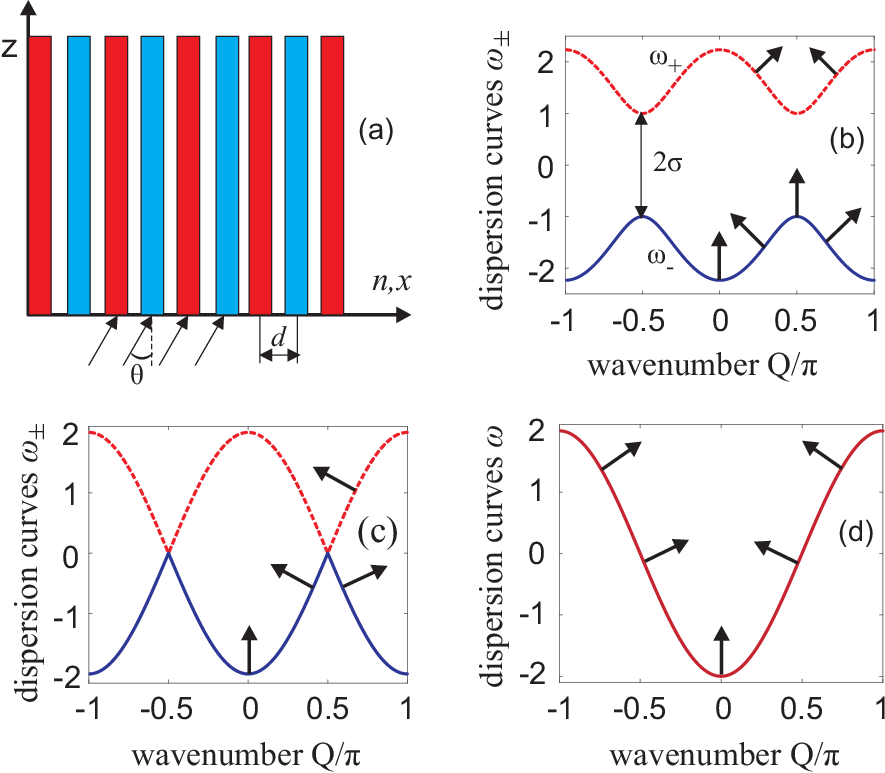}
  \caption{\small{(Color online) (a) Illustrative scheme of a BWA. (b) Dispersion curves $\omega_{\pm}$ for BWAs when $\sigma$ = 1. (c) The same as (b), but now $\sigma$ = 0, i.e., for conventional WAs by using Eqs. (\ref{dispersionBWA}). (d) Dispersion curves $\omega$ for WAs when using Eq. (\ref{dispersionWA}). Arrows in (b-d) illustrate the refraction directions of beams in real space. Other parameters for (b,c,d): $\kappa$ = 1 and $\gamma$ = 0.}}
  \label{fig1}
\end{figure}

The dispersion relationship of Eqs. (\ref{CWCM}) can be found for the linear case and in the absence of the linear potential, i.e., when $\gamma = 0$ and $\alpha$ = 0. In this case, one can insert a plane wave form
\eq{planewave}{a_{n}(Q) \sim \textrm{exp}{[i(Qn - \omega z)]},}
into Eqs. (\ref{CWCM}) and get the following dispersion relationship \cite{sukhorukov1}:
\eq{dispersionBWA}{\omega_{\pm}(Q) =  \pm \sqrt{\sigma^{2} + 4\kappa^{2} \textrm{cos}^{2}Q},}
where the dimensionless quantity $Q$ is the normalized transverse wavenumber. For the sake of brevity, we will refer to quantity $Q$ just as the wavenumber in the rest of this work. This wavenumber also represents the phase difference between optical signals in two adjacent waveguides. Therefore, the wavenumber $Q$ is directly related to the angle $\theta$ between the longitudinal $z-$ axis of the BWA and the beam propagation direction shown in Fig. \ref{fig1}(a). Parameter $\omega$ in Eq. (\ref{planewave}) is the longitudinal wavenumber and is often interpreted as the energy (or frequency) of the plane wave when BWAs are used to simulate relativistic quantum effects governed by the Dirac equation because in that case one needs to use the formal transformation $z \rightarrow t$ with $t$ being the time variable in the Dirac equation \cite{kleinLonghi}.

The two dispersion curves $\omega_{\pm}$ are plotted in Fig. \ref{fig1}(b) with $\sigma$ = 1 and $\kappa$ = 1. The upper dashed red curve in Fig. \ref{fig1}(b) presents the positive-energy state $\omega_{+}$, whereas the lower solid blue curve in Fig. \ref{fig1}(b) presents the negative-energy state $\omega_{-}$. These two minibands get closest to each other at the Dirac points where the wavenumber $Q = \pm \pi/2$. Naturally, around the Dirac points the Zener tunneling between two states are most likely. The gap between these two minibands is $2\sigma$ as shown in Fig. \ref{fig1}(b). In the case of conventional WAs consisting of identical waveguides, i.e., when $\sigma$ = 0, this gap is closed as shown in Fig. \ref{fig1}(c). The arrows in Figs. \ref{fig1}(b-d) illustrate the direction of beams at the corresponding states (see Ref. \cite{dreisowPP} for more details).

Now we want to re-introduce the main result for the beam wavenumber evolution in conventional WAs within the tight-binding approximation which has been obtained earlier in Ref. \cite{tranresonant2}. For conventional WAs in the linear regime, if we insert the plane wave solution in the form of Eq. (\ref{planewave}) into the coupled-mode Eqs. (\ref{CWCM}) with $\sigma$ = 0, one can obtain the following dispersion relationship \cite{jones,lederer}:
\eq{dispersionWA}{\omega(Q) =  -2 \kappa \textrm{cos}Q.}
Note that we have the sign (-) on the right-hand side of Eq. (\ref{dispersionWA}) because we use the sign (-) for $\omega z$ in Eq. (\ref{planewave}). Otherwise, Eq. (\ref{dispersionWA}) will be the same as Eq. (2.5) in Ref. \cite{lederer} [after dropping the trivial term $\beta(\omega)$] and Eq. (2) in Ref. \cite{tranresonant2}.

The dispersion curve described by Eq. (\ref{dispersionWA}) is plotted in Fig. \ref{fig1}(d). It is interesting that dispersion curves shown in Figs. \ref{fig1}(c,d) are both for conventional WAs: Fig. \ref{fig1}(c) is obtained by treating WAs as a special case of BWAs and putting $\sigma$ = 0 in Eqs. (\ref{dispersionBWA}), whereas Fig. \ref{fig1}(d) comes directly from the standard dispersion relationship in the form of Eq. (\ref{dispersionWA}) obtained by eliminating the binary term represented by $\sigma$ in Eqs. (\ref{CWCM}). These two different approaches lead to two totally different interpretations for the BOs (or BZOs) in WAs. Note that the dispersion curve in Fig. \ref{fig1}(d) can be thought of as the combination of the central solid blue curve in Fig. \ref{fig1}(c) in the interval $Q \in [-\pi/2; \pi/2]$ with the two fragments of the upper dashed red curve in Fig. \ref{fig1}(c) in the interval $Q \in [-\pi; -\pi/2]$ and $Q \in [\pi/2; \pi]$.

Following the standard approach detailed in Refs. \cite{lederer,tranresonant2}, after using the Taylor expansion of the dispersion relationship (\ref{dispersionWA}) around the input wavenumber $Q_{0}$, treating the discrete variable $n$ as the continuous one (this approximation is justified if beams are large enough), and introducing the amplitude function $\Psi(n,z) = a_{n,z}\textrm{exp}(-iQ_{0}n)$, one can obtain the following equation from Eqs. (\ref{CWCM}) for conventional WAs (when $\sigma$ = 0):

\eq{NLSE}{\left[ i\de_{z} +\alpha n - \frac{D_{2}}{2}\de_{n}^{2} + \sum_{m\geq3}\frac{D_{m}}{m!}(-i\de_{n})^{m} +  \gamma|\Psi|^{2}\right]\Psi=0,}
where $D_{m} \equiv (d^{m}\omega/dQ^{m})|_{Q_{0}}$ is the $m$-th order diffractive Taylor coefficient.

Without the linear potential term (when $\alpha$ = 0), Eq. (\ref{NLSE}) is similar to the well-known generalized nonlinear Schr\"odinger equation (GNLSE) describing the evolution of light pulses in a single optical fiber \cite{agrawal2} with the only difference that in Eq. (\ref{NLSE}) we have the spatial variable $n$ which performs the role of the temporal variable $t$ in GNLSE. Equation (\ref{NLSE}) is the same as Eq. 4 in Ref. \cite{tranresonant2}.

By using the moment method developed to calculate the rate of soliton self-frequency shift for light pulses in optical fibers \cite{agrawal2,santhanam}, the wavenumber evolution in WAs can be calculated as follows \cite{tranresonant2}:
\eq{SR}{Q(z) = \frac{-i}{2E}\int_{-\infty}^{+\infty} \left(\Psi^{*}\de_{n}\Psi - \Psi\de_{n}\Psi^{*}\right)dn,} where $E = \int_{-\infty}^{+\infty}|\Psi|^{2}dn$ is the total power of the beam light. From Eq. (\ref{NLSE}) and Eq.(\ref{SR}) the law describing the evolution of the central wavenumber of beams in WAs in the linear regime has been analytically found as follows \cite{tranresonant2}:
\eq{fSRWA}{Q(z) = Q_{0} + \alpha z.}
In the rest of this paper, for convenience we fix $\alpha$ = $\pi/50$. As clearly seen from Eq. (\ref{fSRWA}), the wavenumber evolution does not depend on the diffractive coefficients $D_{m}$ at all.

It turns out that the wavenumber evolution described by formula (\ref{fSR}) is also correct with BWAs in the linear regime. In this case, one can replace the linear term $(-1)^{n}\sigma a_{n}$ in Eqs. (\ref{CWCM}) by $\textrm{cos}(\pi n)\sigma a_{n}$ such that we can treat variable $n$ as a continuous one. By doing so, we have to add a new term $\sigma \textrm{cos}(\pi n) \Psi(n,z)$ in Eq. (\ref{NLSE}) as follows:

\eq{NLSEbinary}{\left[ i\de_{z} +\alpha n - \sigma \textrm{cos}(\pi n) - \frac{D_{2}}{2}\de_{n}^{2} + \sum_{m\geq3}\frac{D_{m}}{m!}(-i\de_{n})^{m} +  \gamma|\Psi|^{2}\right]\Psi=0.}

Of course, in Eq. (\ref{NLSEbinary}) the diffractive coefficients $D_{m}$ are calculated by using dispersion relationship (\ref{dispersionWA}), but they do not play any role at all in the evolution of the beam wavenumber, just like what happens with Eq. (\ref{NLSE}). Using the same moment method for Eq. (\ref{NLSEbinary}), we first take the derivative $dQ/dz$ from Eq. (\ref{SR}), then we insert $\de_{z}\Psi$ and $\de_{z}\Psi^{*}$ from Eq. (\ref{NLSEbinary}) to calculate $dQ/dz$. Finally, we obtain the same analytical expression as Eq. (\ref{fSRWA}) for the evolution of beams in BWAs when the wavenumber shift rate is also exactly equal to $\alpha$ - the strength of the linear potential. Moreover, due to the band structure of BWAs shown in Fig. \ref{fig1}(b), after the Zener tunneling takes place, as demonstrated later, another wavenumber spectrum $(Q_{2})$ is generated with a spacing being equal to $\pi$ as compared to the first spectrum $(Q_{1})$ started from $Q_{0}$. As a result, the two central wavenumbers  of beams in BWAs can be explicitly described as follows [see Fig. \ref{fig3}(b)]:

\eq{fSR}{Q_{1}(z) = Q_{0} + \alpha z,}

\eq{fSRweak}{Q_{2}(z) = Q_{0} - \pi + \alpha z.}

For BWAs, as seen in Fig. \ref{fig1}(b), the gap between two minibands is minimum at the Dirac points where $Q_{1} = (m-0.5) \pi$ with $m$ being any integer number. Therefore, the Zener tunneling between two minibands is most favorable at the Dirac points corresponding to the following propagation distances which can easily be obtained by using Eq. (\ref{fSR}):

\eq{ZTdistance}{z_{Tm} = \frac{(m-0.5)\pi - Q_{0}}{\alpha},}
with $m$ = 1, 2, 3, ...

When the central wavenumber $Q_{1}$ reaches the value equal to $\pi$, due to the folding effect reported in Ref. \cite{tranresonant1} in WAs and BWAs and shown later, the wavenumber gets back to the value $-\pi$ and evolves further. At the same time, the other central wavenumber $Q_{2}$ crosses the value 0 and changes its sign. As a result, both beams reach their turning points. Therefore, in the general case, the distances where beams reach the turning points in WAs and BWAs can be easily calculated from Eq. (\ref{fSR}) as follows:

\eq{Zturningpoint}{z_{m} = \frac{m\pi-Q_{0}}{\alpha},}
with $m$ = 1, 2, 3, ...

Formulas (\ref{fSR}) - (\ref{Zturningpoint}) are the main theoretical results of this work which provide a simple analytical expression for the beam wavenumber, the distances where the Zener tunneling can happen and distances where beams reach their turning points during propagation in BWAs depending on the strength of the linear potential $\alpha$ and the central wavenumber $Q_{0}$ of input beams. In Section IV, we show that these analytical results are in perfect agreement with results obtained by simulations of the beam propagation in BWAs in the linear regime.

\section{III. BLOCH-ZENER OSCILLATIONS IN WAVEGUIDE ARRAYS: TWO INTERPRETATIONS}
\label{LBWA}

\begin{figure}[htb]
  \centering \includegraphics[width=0.45\textwidth]{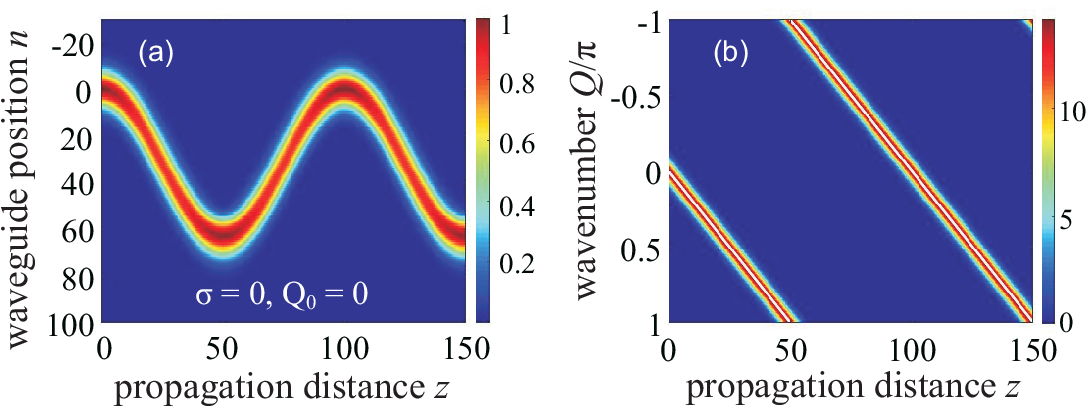}
\caption{\small{(Color online) (a,b) Bloch oscillations of a Gaussian beam in conventional WAs in the $(n,z)$-plane and $(Q,z)$-plane, respectively. Parameters: $\sigma$ = 0, $\alpha = \pi/50$, $Q_{0} = 0$, $\gamma = 0$, $\kappa$ = 1}, and $f$ = 1.}
  \label{fig2}
\end{figure}

As an example to show the validity of Eq. (\ref{fSRWA}) for beams propagating in conventional WAs, we launch a Gaussian beam into the system with the input wavenumber $Q_{0}$ = 0 (normal incidence). In the rest of this work, we use Gaussian input beams in the following form:

\eq{gaussianinitial}{a_{n} = f\textrm{exp}\left(-\frac{n^{2}}{w_{0}^2}\right)\textrm{exp}(iQ_{0}n),}
where $f$ is a factor representing the peak amplitude of Gaussian input beams, $w_{0}$ characterizes the width of Gaussian input beams which is fixed at $w_{0}$ = 8.0 in this work. Using this Gaussian input beam as the initial condition to numerically integrate Eqs. (\ref{CWCM}) in the linear regime and without the binary term ($\gamma$ = 0 and $\sigma$ = 0) with the help of the Runge-Kutta fourth order, we obtain the beam dynamics in the $(n,z)$-plane in Fig. \ref{fig2}(a) which shows the evolution of the quantity $|a_n(z)|$ during beam propagation in the WA. Then we take the Fourier transform of $a_{n}(z)$ with respect to the transverse coordinate $n$ and show the evolution of the beam wavenumber in the $(Q,z)$-plane in Fig. \ref{fig2}(b) where the solid white line represents the beam evolution based on Eq. (\ref{fSR}). As mentioned above, because the WA is transversally periodic, the wavenumber $Q$ only evolves in the first Brillouin zone in the interval $[-\pi;\pi]$. Therefore, one can observe the folding effect reported in Ref. \cite{tranresonant1}: when the wavenumber $Q$ reaches the value $Q = \pi$, it folds back into the first Brillouin zone and $Q$ continues from the value $Q = -\pi$ as shown in Fig. \ref{fig2}(b). Note that, unlike the artificial manipulation of the solid white line at the edges of the first Brillouin zone, the spectrum evolution of the beam shown in Fig. \ref{fig2}(b) is quite natural without any artificial intervention.

As clearly demonstrated in Fig. \ref{fig2}(b), Eq. (\ref{fSR}) perfectly agrees with the simulation results. As the input Gaussian beam has a finite width ($w_{0} = 8.0$), its spectrum also has a finite width and evolves during propagation as accurately described by Eq. (\ref{fSR}). One direct consequence of the wavenumber shift in the linear regime is the Bloch oscillations \cite{pertsch,lenz,ghulinyan,trompeter} and demonstrated in Fig. \ref{fig2}. We can interpret Bloch oscillations based on Eq. (\ref{fSR}) in the following way. In Fig. \ref{fig2}(a) a Gaussian beam is normally launched into a WA made of $N=241$ waveguides with initial wavenumber $Q_{0}=0$, but because of the wavenumber shift due to the linear potential, its wavenumber increases linearly and remains positive, thus the beam propagates downwards in Fig. \ref{fig2}(a) towards positive values of $n$. At the distance $z = (\pi - Q_{0})/\alpha = 50$, the wavenumber reaches the value $Q = \pi$, and the beam is parallel to the $z$-axis of the WA as the input beam [see also arrows in Fig. \ref{fig1}(d)]. However, due to the folding of the spectrum, the wavenumber folds back into the first Brillouin zone, and takes the value $Q = -\pi$ and becomes negative. This means that the beam must change its direction, and starts moving upwards in Fig. \ref{fig2}(a) towards negative values of $n$. This process is repeated continuously in a sinusoidal fashion with a period $z_{B} = 2\pi/\alpha = 100$. This oscillation is none other than the Bloch oscillation with the period $z_{B} = 2\pi/\alpha$ having been found earlier with a different approach based on the Wannier-Stark ladders \cite{pertsch,lenz,ghulinyan,trompeter}. Therefore, if we use the band structure of WAs shown in Fig. \ref{fig2}(b), we have the pure Bloch oscillations without any Zener tunneling, because there is just one band in Fig. \ref{fig1}(d). This interpretation of BOs has been provided in Ref. \cite{tranresonant2}.

However, if we use the band structure shown in Fig. \ref{fig1}(c) for the WA as a special case of BWAs with $\sigma = 0$, then we have a totally different interpretation where the Zener tunneling is complete $(100/\%)$ \cite{dreisowZener}. With $Q_{0}$ = 0 and $\alpha = \pi/50$ we have the distances where the Zener tunneling takes place as follows as described by Eq. (\ref{ZTdistance}): $z_{T1} = 25 = z_{B}/4$, $z_{T2} = 75 = 3z_{B}/4$, $z_{T3} = 125 = 5z_{B}/4$. At the distance $z_{T1} = 25$, the central wavenumber of the beam reaches the Dirac point $Q = \pi/2$ which is the right tip of the lower solid blue curve in Fig. \ref{fig1}(c). Because the gap between two minibands in Fig. \ref{fig1}(c) is equal to zero ($2\sigma$ = 0) in WAs, and because the beam does not change its direction in the vicinity of the distance $z_{T1} = 25$ as shown in Fig. \ref{fig2}(a), the beam must switch to the upper dashed curve in Fig. \ref{fig1}(c). Therefore, one can say that the Zener tunneling happens at $z_{T1} = 25$ and it is complete, i.e., all of its energy is tunneled to the upper state. At the distance $z$ = 50 in Fig. \ref{fig2}(b), the beam wavenumber reaches the edge value $Q = \pi$, thus it has to fold back to get another edge value $Q = -\pi$ while the beam is still on the upper state shown in Fig. \ref{fig1}(c). After getting the value $Q = -\pi$, the beam must change its direction as clearly shown in Fig. \ref{fig1}(c) and confirmed by Fig. \ref{fig2}(a). At the distance $z_{T2} = 75$, the wavenumber again reaches another Dirac point with $Q = -\pi/2$, and because the beam does not change its direction at this point as shown in Fig. \ref{fig2}(a), it must completely switch to the lower miniband in Fig. \ref{fig1}(c). This means that at the distance $z_{T2} = 75$, the Zener tunneling happens again and transfers all of the beam energy to the lower miniband. Afterwards, the beam continues to evolve towards the value $Q = 0$ at the distance $z_{B} = 100$ and complete one period of the Bloch-Zener oscillation.

\section{IV. BLOCH-ZENER OSCILLATIONS IN LINEAR BINARY WAVEGUIDE ARRAYS}
\label{LBWA}

\begin{figure}[htb]
  \centering \includegraphics[width=0.45\textwidth]{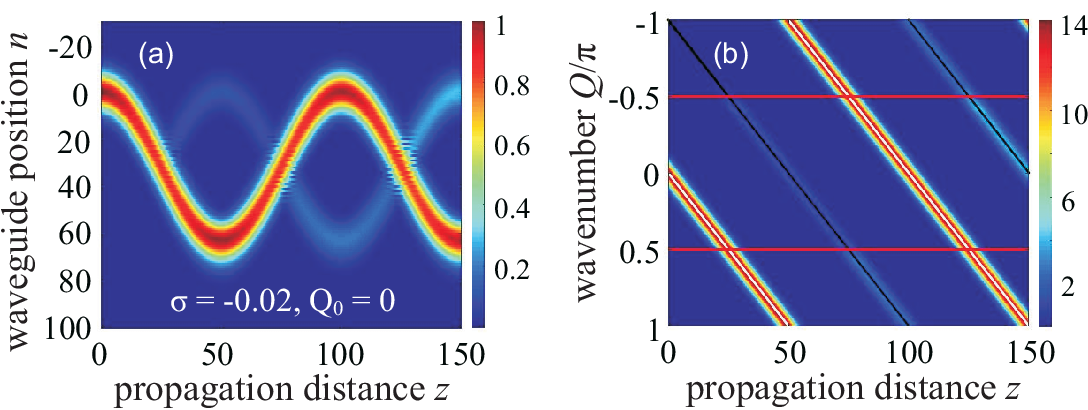}
\caption{\small{(Color online) (a,b) Bloch-Zener oscillations of a Gaussian beam in a BWA in the $(n,z)$-plane and $(Q,z)$-plane, respectively. Parameters: $\sigma = - 0.02$, $\alpha = \pi/50$, $Q_{0} = 0$, $\gamma = 0$, and $f$ = 1}.}
  \label{fig3}
\end{figure}

Now it is time for us to study the Bloch-Zener oscillations in BWAs in the linear regime under the action of an external linear potential. First, we open the gap between two minibands just a little bit, i.e., when the Dirac mass $\sigma$ is small. In Fig. \ref{fig3} we normally launch the Gaussian beam into the BWA with $\sigma = -0.02$. The evolution of beams in the $(n,z)$-plane is shown in Fig. \ref{fig3}(a), whereas the evolution of the beam wavenumber in the $(Q,z)$-plane is shown in Fig. \ref{fig3}(b). At first, the beam moves downwards as shown in Fig. \ref{fig3}(a), and the beam wavenumber shown in Fig. \ref{fig3}(b) linearly increases along the solid white line described by Eq. (\ref{fSR}). At the distance $z_{T1} = 25 = z_{B}/4$, the wavenumber reaches the Dirac point $Q = \pi/2$ and the Zener tunneling takes place. Indeed, as shown in Fig. \ref{fig3}(a), most of the beam power ($99\%$) continues to move downwards which is also in agreement with Fig. \ref{fig3}(b) where most of the beam spectrum continues to develop along the solid white line described by Eq. (\ref{fSR}). This means that most of the beam power is switched to the upper miniband after crossing the Dirac point $Q = \pi/2$. However, one can see a weak beam is generated and moves upwards in Fig. \ref{fig3}(a). Specifically, it takes the symmetrical direction with respect to the $z$-axis as compared to the intense beam moving downwards. This weak beam still stays on the lower miniband and now its central wavenumber starts from the other Dirac point with $Q = -\pi/2$ because the weak beam now has the motion direction symmetrical to the intense beam direction. The central wavenumber of the weak beam also linearly increases with the same rate $\alpha$ as the intense beam. Its evolution is described by Eq. (\ref{fSRweak}) and is plotted by the solid black lines in Fig. \ref{fig3}(b). The spacing between two central wavenumbers of two beams in wavenumbers is $\pi$ as clearly confirmed by Fig. \ref{fig3}(b) in which we also mark the two Dirac points at $Q = \pm \pi/2$ by two horizontal solid red lines.

As clearly shown in Fig. \ref{fig3}(b), the simulation results for the evolution of the two central wavenumbers of beams in BWAs due to the Bloch oscillations, Zener tunneling, and folding effect are in perfect agreement with two simple analytical formulas Eq. (\ref{fSR}) and Eq. (\ref{fSRweak}) presented by the solid white line and solid black one, respectively. At the distance $z = 50$, the folding effect at the edge of the Brillouin zone takes place with the intense beam, whereas the central wavenumber of the weak beam reaches the value $Q_{2} = 0$. As a result, both beams turn back.

At the distance $z_{T2} = 75 = 3z_{B}/4$ in Fig. \ref{fig3}, both beams reach Dirac points at $Q = \pm \pi/2$, and the Zener tunneling happens again with them: the intense beam now is switched to the lower miniband $\omega_{-}$, whereas the weak beam is switched to the upper miniband $\omega_{+}$.

At the distance $z_{B} = 2\pi/\alpha = 100$ in Fig. \ref{fig3}, the folding effect at the edge of the Brillouin zone takes place with the weak beam, whereas the central wavenumber of the intense beam now reaches the value $Q_{1} = 0$. As a result, both beams also turn back. This combination of Bloch oscillations and Zener tunneling repeats again and again as beams continue to propagate further.

\begin{figure}[htb]
  \centering \includegraphics[width=0.45\textwidth]{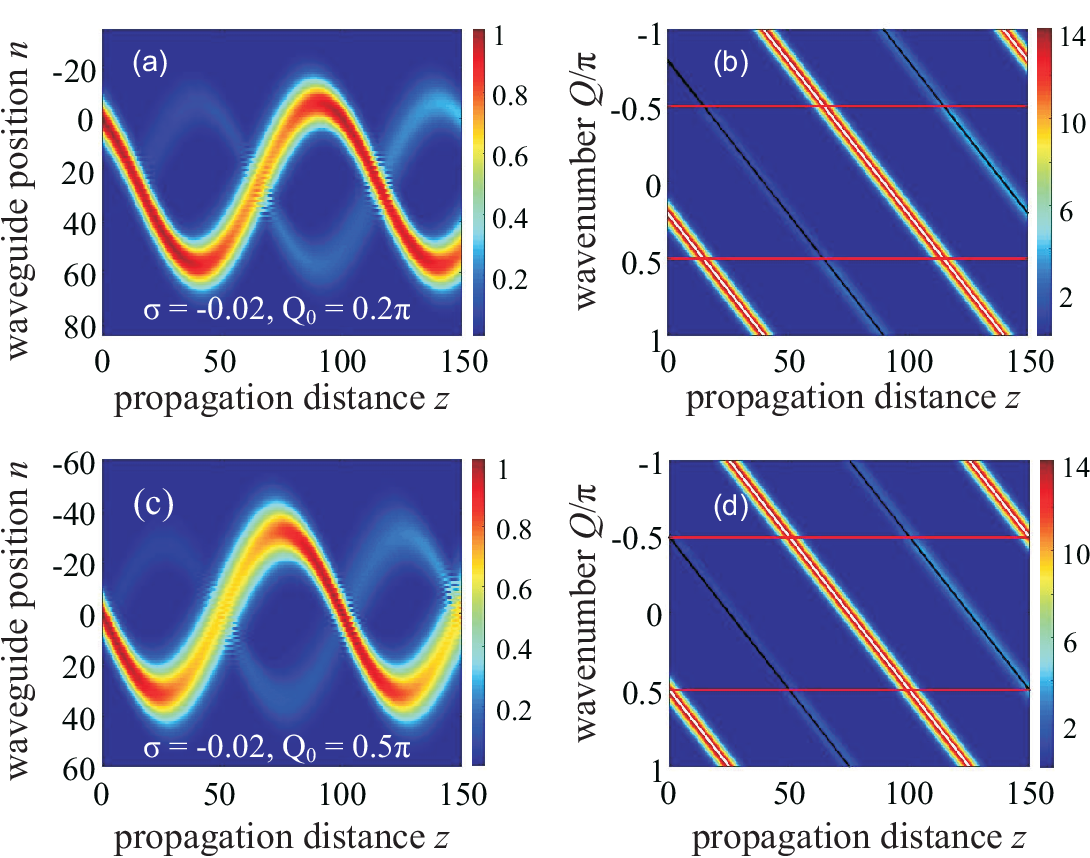}
\caption{\small{(Color online) (a,b) Bloch-Zener oscillations of a Gaussian beam in a BWA in the $(n,z)$-plane and $(Q,z)$-plane, respectively, when $Q_{0} = 0.2\pi$. (c,d) The same as (a,b), but now $Q_{0} = 0.5\pi$. Other parameters: $\sigma - 0.02$, $\alpha = \pi/50$, $\gamma = 0$, and $f$ = 1}.}
  \label{fig4}
\end{figure}

In Figs. \ref{fig4}(a,b), we launch the Gaussian beam obliquely into the system such that $Q_{0} = 0.2\pi$ instead of $Q_{0} = 0$ in Fig. \ref{fig3}. All other parameters in Fig. \ref{fig4}(a,b) are the same as in Fig. \ref{fig3}.

All the Bloch-Zener oscillation scenarios in Figs. \ref{fig4}(a,b) are practically the same as in Fig. \ref{fig3}. With the input wavenumber now $Q_{0} = 0.2\pi$ one can easily calculate the distances where the Zener tunneling takes places based on Eq. (\ref{ZTdistance}): $z_{T1}$ = 15; $z_{T2}$ = 65; $z_{T3}$ = 115; and the distances where both beams reach the turning points based on Eq. (\ref{Zturningpoint}): $z_{1}$ = 40; $z_{2}$ = 90; $z_{3}$ = 140. All these special distances are perfectly confirmed by beam evolution simulations shown in Figs. \ref{fig4}(a,b). At the distance where the Zener tunneling takes place first $z_{T1}$ = 15, $99 \%$ of the beam power is tunneled to the upper miniband, just the same as in Fig. \ref{fig3}.

In Figs. \ref{fig4}(c,d), we launch the Gaussian beam obliquely at the Dirac point into the system such that $Q_{0} = 0.5\pi$. All other parameters in Fig. \ref{fig4}(c,d) are the same as in Fig. \ref{fig3}. As expected, the Zener tunneling takes place right at the beginning in which $99.6 \%$ of the beam power is switched to the upper miniband. Other distances where the Zener tunneling takes place in Figs. \ref{fig4}(c,d) can be calculated directly from Eq. (\ref{ZTdistance}): $z_{T2}$ = 50; $z_{T3}$ = 100; $z_{T4}$ = 150. The distances where both beams reach the turning points based on Eq. (\ref{Zturningpoint}): $z_{1}$ = 25; $z_{2}$ = 75; $z_{3}$ = 125.

\begin{figure}[htb]
  \centering \includegraphics[width=0.45\textwidth]{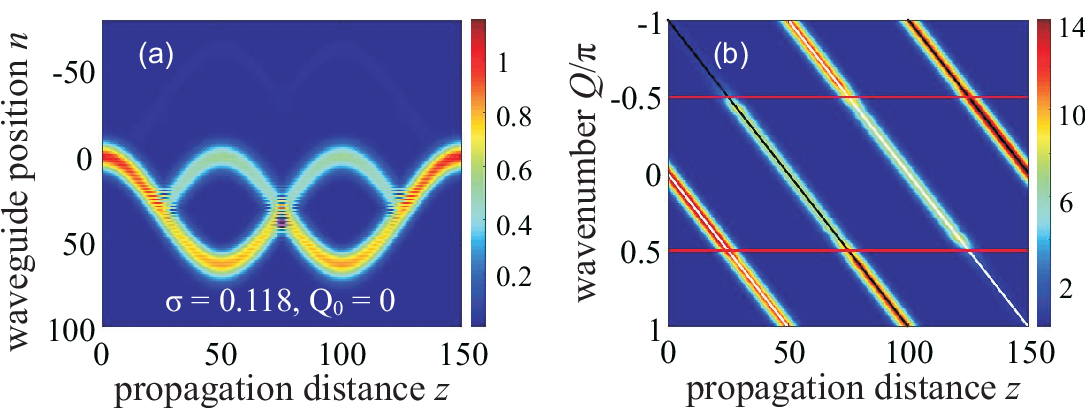}
\caption{\small{(Color online) (a,b) Bloch-Zener oscillations of a Gaussian beam in a BWA in the $(n,z)$-plane and $(Q,z)$-plane, respectively. Parameters: $\sigma = 0.118$, $\alpha = \pi/50$, $Q_{0} = 0$, $\gamma = 0$, and $f$ = 1}.}
  \label{fig5}
\end{figure}

As shown in Fig. \ref{fig3} and Fig. \ref{fig4}, the wavenumber evolutions of beams are perfectly described by Eq. (\ref{fSR}) and Eq. (\ref{fSRweak}) when the Dirac mass is quite small ($\sigma = -0.02$). Our simulations with much larger values of $\sigma$ (not included here) also confirm the validity of Eq. (\ref{fSR}) and Eq. (\ref{fSRweak}).

Note that one can tune parameters so that one beam ceases to exist at some of distances described by Eq. (\ref{ZTdistance}) with $m \geq 2$. This is because at these distances, both beams undergo the Zener tunnelings which altogether can lead to the vanishing of one beam \cite{breid}. In Fig. \ref{fig5} we tune $\sigma$ to the value 0.118 and launch the Gaussian beam normally ($Q_{0} = 0$) into the BWA. In this case, one beam stops to exist at the distance $z_{T3}$ = 75. Note also that in Fig. \ref{fig5} the fraction of the beam power switched to the upper miniband at the distance $z_{T1}$ = 25 is $70.6 \%$ which is smaller than in Fig. \ref{fig3} and Fig. \ref{fig4} because the gap between the two minibands is increased now. The evolution of the beam wavenumber shown in Fig. \ref{fig5}(b) also well confirms the validity of Eq. (\ref{fSR}) and Eq. (\ref{fSRweak}).

\section{V. BLOCH-ZENER OSCILLATIONS IN NONLINEAR BINARY WAVEGUIDE ARRAYS}
\label{NLBWA}

So far in this work we have just investigated the beam dynamics in WAs and BWAs in the linear regime when $\gamma = 0$. In this Section we investigate the influence of the nonlinearity on the BZOs in BWAs. It has been reported in Ref. \cite{morandotti} that the nonlinearity leads to the negative influence on BOs in WAs such as the loss of beam recovery, symmetry breaking and beam spreading. It has been also shown in Ref. \cite{tranresonant2} that the strong nonlinearity can completely destroy the BOs in WAs. Our simulations (not included here) also show that when the nonlinearity is strong. i.e., when the factor $f$ is large enough while fixing $\gamma$, the BZOs are also destroyed in BWAs where beams strongly spread out. However, as expected, if the nonlinearity is not so strong, one still can observe BZOs and the wavenumber evolution can still be described by Eq. (\ref{fSR}) and Eq. (\ref{fSRweak}).

\begin{figure}[htb]
  \centering \includegraphics[width=0.45\textwidth]{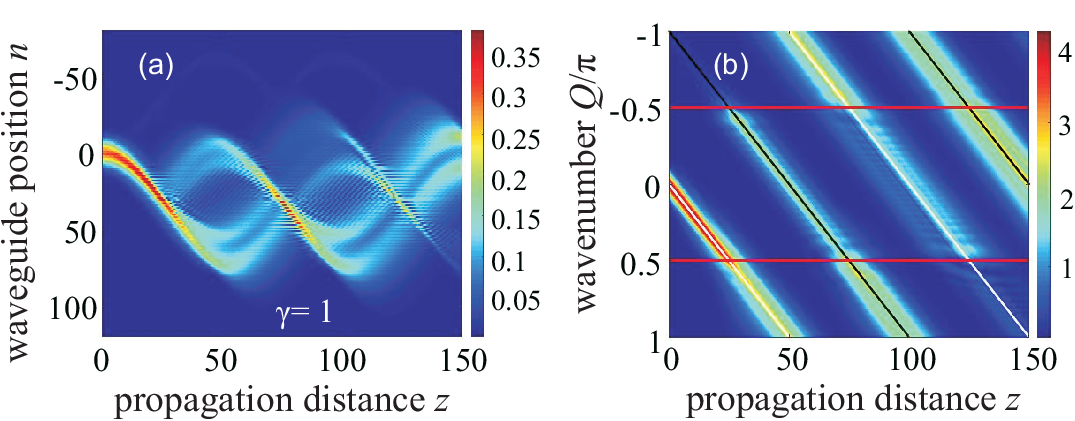}
\caption{\small{(Color online) (a,b) Bloch-Zener oscillations of a Gaussian beam in a nonlinear BWA in the $(n,z)$-plane and $(Q,z)$-plane, respectively. Parameters: $\sigma = 0.118$, $\alpha = \pi/50$, $Q_{0} = 0$, $\gamma = 1$, and $f$ = 0.3}.}
  \label{fig6}
\end{figure}

As an example, in Figs. \ref{fig6}(a,b) we show the beam evolution and wavenumber evolution, respectively, when we set $f$ = 0.3 and $\gamma$ = 1. All other parameters are the same as in Fig. \ref{fig5}. Note that because Fig. \ref{fig5} is obtained in the linear regime, the beam dynamics does not change at all if $f$ takes any value different from zero. Although the spreading of beams is evident in Fig. \ref{fig6}(a) due to the nonlinearity, the central wavenumber of beams still quite well obeys Eq. (\ref{fSR}) and Eq. (\ref{fSRweak}) represented by the solid white line and solid black line in Fig. \ref{fig6}(b), respectively.

\section{VI. CONCLUSIONS}
\label{conclusions}

We have investigated the Bloch-Zener oscillations in WAs and BWAs under action of the linear potential by using an approach based on the wavenumber evolution. In the linear regime, the simple analytical expressions for the central wavenumber of beams in BWAs have been found and are in perfect agreement with simulations results. The beam wavenumber linearly depends on the gradient of the average refractive index of waveguides in the transverse direction and also depends on the initial wavenumber of input beams. Thanks to these simple laws for wavenumber evolutions, one can easily find the distances where the Zener tunneling happens and the distances where beams change their motion directions. In the case of WAs, only one beam exists and periodically oscillates whose dynamics can be interpreted by two totally different ways: either as pure Bloch oscillations or a combination of Bloch oscillations with a complete Zener tunneling between two gapless minibands at certain distances. In the case of BWAs, after the first Zener tunneling at the Dirac point, two beams with wavenumbers separated by $\pi$ are formed and propagate in symmetrical directions with respect to the $z$-axis of BWAs. The nonlinearity of Kerr type in the system can destroy the Bloch-Zener oscillations of beams. However, if the nonlinearity is not so strong, the Bloch-Zener oscillations of beams are maintained quite well, and the central wavenumber of beams also evolves linearly.

\end{document}